\documentclass[12pt]{article}
\usepackage{graphicx}
\usepackage{amssymb}

\newcommand{\beq}{\begin{equation}}
\newcommand{\eeq}{\end{equation}}

\newcommand{\n}{\mbox{${\mathbf n}$}}
\newcommand{\p}{\mbox{${\mathbf p}$}}
\newcommand{\q}{\mbox{${\mathbf q}$}}
\newcommand{\s}{\mbox{${\mathbf s}$}}

\newcommand{\bl}{\mbox{${\mathbf l}$}}

\newcommand{\br}{\mbox{${\mathbf r}$}}
\newcommand{\bS}{\mbox{${\mathbf S}$}}

\newcommand{\bv}{\mbox{${\mathbf v}$}}

\newcommand{\vom}{\mbox{{\boldmath$\omega$}}}

\newcommand{\vrho}{\mbox{{\boldmath$\rho$}}}

\newcommand{\ga}{\mbox{${\gamma}$}}

\newcommand{\de}{\mbox{${\delta}$}}

\newcommand{\ka}{\mbox{${\varkappa}$}}
\newcommand{\la}{\mbox{${\lambda}$}}
\newcommand{\ep}{\mbox{${\varepsilon}$}}
\newcommand{\om}{\mbox{${\omega}$}}

\newcommand{\pa}{\mbox{${\partial}$}}

\begin{document}

\begin{titlepage}

\begin{center}

{\bf \large Quantum long-range interactions in general relativity}

\vspace{1cm}

I.B. Khriplovich\footnote{khriplovich@inp.nsk.su} and G.G.
Kirilin\footnote{g\_kirilin@mail.ru}

\vspace{1cm}

Budker Institute of Nuclear Physics\\
630090 Novosibirsk, Russia\\
and Novosibirsk University

\end{center}

\bigskip

\begin{abstract}

We consider one-loop effects in general relativity that result in
quantum long-range corrections to the Newton law, as well as to
the gravitational spin-dependent and velocity-dependent
interactions. Some contributions to these effects can be
interpreted as quantum corrections to the Schwarzschild and Kerr
metrics.

\bigskip


\end{abstract}

\vspace{8cm}

\end{titlepage}

\section{Introduction}

It has been recognized long ago that quantum effects in general
relativity can generate long-range corrections to the Newton law.
Such corrections due to the contribution by photons and massless
neutrinos to the graviton polarization operator were calculated by
Radkowski~\cite{rad}, Capper, Duff, and Halpern \cite{caph},
Capper and Duff \cite{cap}, Duff and Liu \cite{dli}. The
corresponding quantum correction to the Newton potential between
two bodies with masses $m_1$ and $m_2$ is
\beq\label{a1}
U_{\gamma\nu}=-\,\frac{4+N_\nu}{15\pi}\,\frac{k^2\hbar m_1
m_2}{c^3 r^3}\,,
\eeq
where $N_\nu$ is the number of massless two-component neutrinos,
$k$ is the Newton gravitational constant.

The reason why the problem allows a closed solution is as follows.
The Fourier-transform of $1/r^3$ is
\beq\label{fur}
\int d\br \,\frac{\exp (-i \q \br)}{r^3}\,= -\,2\pi \ln q^2.
\eeq
This singularity in the momentum transfer $\q$ implies that the
discussed correction can be generated only by diagrams with two
massless particles in the $t$-channel. The number of such diagrams
of second order in $k$ is finite, and their logarithmic part in
$q^2$ can be calculated unambiguously.

\begin{figure}[h]
\begin{center}
\includegraphics*{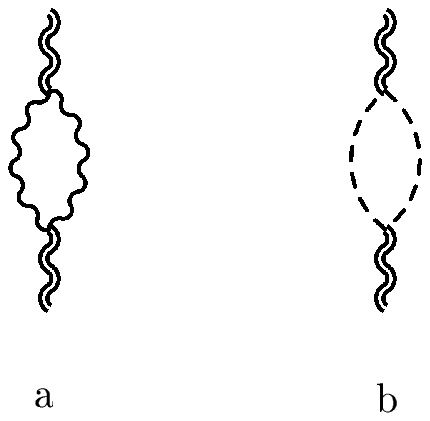}\hfill
\caption{Graviton loop}\label{loop}
\end{center}
\end{figure}

The analogous diagrams with gravitons and ghosts in the loop,
Figs. \ref{loop}a,b, were considered in Refs.
\cite{rad,capl,duf,hoo}. (Here and below, wavy lines refer to
quantum fluctuations of metric, double wavy lines denote a
background gravitational field; dashed lines here refer to
ghosts.)
Clearly, other diagrams with two gravitons in the $t$-channel
contribute as well to the discussed correction $\sim 1/r^3$. This
was pointed out long ago by Boulware and Deser \cite{bode},
together with indicating explicitly all relevant diagrams.

The problem of quantum correction to the Newton law is certainly
interesting from the theoretical point of view. It was addressed
later by Donoghue \cite{don,don1,don2,don3}, Muzinich and
Vokos~\cite{muz}, Akhundov, Belucci, and Shiekh \cite{akh}, as
well as by Hamber and Liu \cite{ham}. Unfortunately, as
demonstrated in~\cite{kk}, neither of these attempts was
satisfactory.

Then the discussed problem was considered quantitatively in our
previous paper \cite{kk}. Therein, all relevant diagrams, except
one (see Fig. \ref{vertex}b below), were calculated correctly. In
a recent paper by Bjerrum-Bohr, Donoghue, and Holstein \cite{bb}
this last diagram is calculated correctly\footnote{Both previous
results for this contribution, by Donoghue~\cite{don1} and by
us~\cite{kk}, were incorrect.}, and our results for all other
contributions are confirmed.

The content of our present work is as follows. Using the
background field technique by 't~Hooft and Veltman \cite{hoo}, we
construct invariant operators that describe quantum power
corrections in general relativity. In the limit when one of the
interacting particles is heavy, one can interpret some of the
derived corrections as quantum corrections to the Schwarzschild
and Kerr metric. Here our results differ essentially from those by
Bjerrum-Bohr, Donoghue, and Holstein~\cite{bbo}.

We demonstrate also in an elementary way that, to our accuracy,
the spin-independent part of the discussed corrections for spinor
particles coincides with the corrections for scalar ones. It
implies in particular that the obtained quantum corrections to the
Schwarzschild metric are universal, i. e. independent of the spin
of the central body. For some loop diagrams relevant to the
problem, the mentioned coincidence of spin-independent
contributions of spinor particles with the corresponding results
for scalar ones was proven previously in \cite{bbo} by direct
calculation.

With the constructed effective operators we not only derive easily
the corrections to the Newton law. They allow us to obtain quantum
corrections to other gravitational effects: spin-dependent and
velocity-dependent interactions. In the present paper we confine
mainly to the case of scalar particles. Therefore, by spin we mean
here the internal angular momentum of a compound particle with
scalar constituents.

We comment also on the problem of the classical relativistic
corrections to the Newton law. Our conclusions here agree
completely with the results by Einstein, Infeld,
Hoffmann~\cite{eih}, Eddington, Clark~\cite{ec}, Iwasaki~\cite{iw}
(see also the textbook~\cite{ll}, \S 106), but on some point we
disagree essentially with the statements by Bjerrum-Bohr,
Donoghue, and Holstein~\cite{bb}.

\section{Propagators and Vertices}

We use below the units with $c=1$, $\hbar=1$. Our metric signature
is $\; {\rm diag}(1, -1,-1,-1)$.

The graviton operator $h_{\mu\nu}$ describes quantum fluctuations
of the metric $g_{\mu\nu}$ in the background metric
$g_{\mu\nu}^o$:
\begin{eqnarray}
g_{\mu\nu}=g_{\mu\nu}^o+\ka\, h_{\mu\nu}\,;  \qquad \ka^2=32\pi
k\,= 32\pi l^2_p\,. \label{l:1}
\end{eqnarray}
We use for $h_{\mu\nu}$ the gauge condition
\beq\label{harm}
h^\mu_{\nu;\mu} - \,\frac{1}{2}\,h^\mu_{\mu;\nu} =0;
\eeq
here indices of $h_{\mu\nu}$ are raised with the background metric
$g_{\mu\nu}^o$, and the covariant derivatives are taken in the
background field $g_{\mu\nu}^o$. The free graviton propagator is
\beq\label{gp}
D_{\mu\nu,\alpha\beta}(q)=i\,\frac{P_{\mu\nu,\alpha\beta}}{q^2+i0}\,;\qquad
P_{\mu\nu,\alpha\beta}\,=\,\frac{1}{2}\,(\delta_{\mu\alpha}\delta_{\nu\beta}
+\delta_{\nu\alpha}\delta_{\mu\beta}-
\delta_{\mu\nu}\delta_{\alpha\beta})\,.
\eeq
The tensor $P_{\mu\nu,\alpha\beta}$ is conveniently presented as
\cite{hoo}
\[
P_{\mu\nu,\alpha\beta}\,=\,I_{\mu\nu,\alpha\beta} -
\,\frac{1}{2}\,\delta_{\mu\nu}\delta_{\alpha\beta}\,,
\]
where
$I_{\mu\nu,\alpha\beta}=\,\frac{1}{2}\,(\delta_{\mu\alpha}\delta_{\nu\beta}
+\delta_{\nu\alpha}\delta_{\mu\beta})$ is a sort of a unit
operator with the property
\[
I_{\mu\nu,\alpha\beta}t_{\alpha\beta}=t_{\mu\nu}
\]
for any symmetric tensor $t_{\alpha\beta}$. Let us note the
following useful identity:
\beq\label{id}
P_{\alpha\beta,\kappa\lambda} P_{\kappa\lambda,\gamma\delta} =
I_{\alpha\beta,\gamma\delta}\,.
\eeq
\begin{figure}[h]
\begin{center}
\includegraphics*{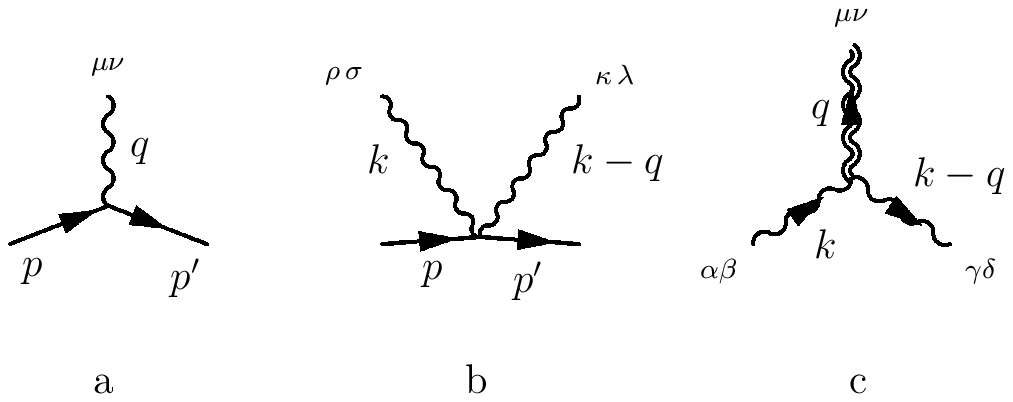}\hfill
\caption{Gravitational vertices}\label{vertices}
\end{center}
\end{figure}

The propagators of scalar and spinor particles are the usual ones:
\[
D(p)=i\frac{1}{p^2-m^2+i0}\,, \quad {\rm and} \quad
G(p)=i\frac{1}{\hat{p} - m + i0}\,,
\]
respectively.

Single-graviton vertex both for scalar and spinor particles (see
Fig. \ref{vertices}a) are related to the energy-momentum tensor
$T_{\alpha\beta}(p,p')$ of the corresponding particle as follows:
\beq
V_{\alpha\beta}(p,p')= -\,i
\,\frac{\ka}{2}\,T_{\alpha\beta}(p,p')\,.
\eeq
The explicit expressions for scalar and spinor particles are
\beq\label{1sc}
V_{\alpha\beta}^{(0)}(p,p')= -\,i \,\frac{\ka}{2}\left[p_\alpha
{p'}_\beta + {p'}_\alpha p_\beta -
\delta_{\alpha\beta}(pp'-m^2)\right],
\eeq
and
\begin{eqnarray}\label{1sp}
V_{\mu\nu}^{(1/2)} = - \frac{i\,\varkappa}{4}\,\bar{u}\,(p')\,[
I_{\mu\nu\alpha\beta}P_\alpha\gamma_\beta -
\delta_{\mu\nu}(\hat{P} - 2 m)]\,u(p)\,,
\end{eqnarray}
respectively; here $P=p+p^{\prime}$.

The contact interaction of a scalar particle with two gravitons
(see Fig. \ref{vertices}b) is
\beq\label{v2}
V_{\varkappa\lambda,\rho\sigma}^{(0)} = i\ka^2\;
[\;I_{\kappa\lambda,\alpha\delta}I_{\delta\beta,\rho\sigma}
(p_{\alpha}p'_{\beta}+p'_{\alpha}p_{\beta})-\,\frac{1}{2}\,
(\delta_{\kappa\lambda}I_{\rho\sigma,\alpha\beta}+\delta_{\rho\sigma}
I_{\kappa\lambda,\alpha\beta}) p_\alpha p'_\beta
\eeq
\[
+\,\frac{(p'-p)^2}{4}\,(I_{\kappa\lambda,\rho\sigma}
-\frac{1}{2}\delta_{\kappa\lambda}\delta_{\rho\sigma})\;].
\]
To our accuracy, one can neglect in this expression the last term,
with $(p'-p)^2=q^2$, and rewrite the vertex conveniently as
\beq
V_{\varkappa\lambda,\rho\sigma}^{(0)} =
i\ka^2\,[\,I_{\varkappa\lambda,\alpha\delta}I_{\delta\beta,\rho\sigma}T_{\alpha\beta}
-\frac{1}{4}(\de_{\varkappa\lambda}T_{\rho\sigma}+\de_{\rho\sigma}T_{\varkappa\lambda})]\,.
\eeq
We use the two-graviton vertices on-mass-shell only. Therefore,
the terms with the Kronecker $\de$ entering the energy-momentum
tensor in the last expression are also proportional to $q^2$, and
thus can be neglected.

The contact two-graviton interaction of a spinor particle (see
Fig. \ref{vertices}b) can be written on-mass-shell as follows:
\[
V^{(1/2)}_{\varkappa\lambda,\,\rho\sigma} =
i\,\frac{\varkappa^2}{8}
\left[\frac{3}{2}\left(I_{\varkappa\lambda,\,\mu\beta}
\,I_{\rho\sigma,\,\beta\alpha}+I_{\rho\sigma,\,\mu\beta}
\,I_{\varkappa\lambda,\,\beta\alpha} \right) P_\mu - \right.\\
\]
\begin{equation}
\left.- \delta_{\varkappa\lambda} I_{\rho\sigma,\,\mu\alpha} P_\mu
- \delta_{\rho\sigma} I_{\varkappa\lambda,\,\mu\alpha} P_\mu
\right] \bar{u}\,(p')\gamma^\alpha u(p) \label{fv2}
\end{equation}
\[
= \,
i\varkappa^2\left[\frac{3}{4}\,I_{\varkappa\lambda,\alpha\delta}I_{\delta\beta,\rho\sigma}T_{\alpha\beta}
-\,\frac{1}{4}(\de_{\varkappa\lambda}T_{\rho\sigma}+\de_{\rho\sigma}T_{\varkappa\lambda})\right].
\]

As to the 3-graviton vertex (see Fig. \ref{vertices}c), which has
the most complicated form, we follow~\cite{hoo,bb} in representing
it as
\begin{eqnarray}\label{v3}
&V_{\mu\nu,\alpha\beta,\gamma\delta}=&-i\,\frac{\ka}{2}\,\sum_i\,
{}^iv_{\mu\nu,\alpha\beta,\gamma\delta};\\ &
{}^1v_{\mu\nu,\alpha\beta,\gamma\delta}=&
P_{\alpha\beta,\gamma\delta}\,[k_\mu k_\nu+(k-q)_\mu
(k-q)_\nu+q_\mu q_\nu-
\frac{3}{2}\,\delta_{\mu\nu}q^2],\nonumber\\ &
{}^2v_{\mu\nu,\alpha\beta,\gamma\delta}=&2 q_\varkappa q_\lambda [
I_{\varkappa\lambda,\alpha\beta}I_{\mu\nu,\gamma\delta}
+I_{\varkappa\lambda,\gamma\delta}I_{\mu\nu,\alpha\beta}-
I_{\varkappa\mu,\alpha\beta}I_{\lambda\nu,\gamma\delta}
-I_{\varkappa\nu,\alpha\beta}I_{\lambda\mu,\gamma\delta}],\nonumber\\
& {}^3v_{\mu\nu,\alpha\beta,\gamma\delta}=& q_{\varkappa} q_{\mu}
(\delta_{\alpha\beta}I_{\varkappa\nu,\gamma\delta}+
\delta_{\gamma\delta}I_{\varkappa\nu,\alpha\beta})+ q_{\varkappa}
q_{\nu} (\delta_{\alpha\beta}I_{\varkappa\mu,\gamma\delta}+
\delta_{\gamma\delta}I_{\varkappa\mu,\alpha\beta})\nonumber\\
&&-q^2(\delta_{\alpha\beta}I_{\mu\nu,\gamma\delta}+
\delta_{\gamma\delta}I_{\mu\nu,\alpha\beta})-
\delta_{\mu\nu}q_\varkappa q_\lambda
(\delta_{\alpha\beta}I_{\gamma\delta,\varkappa\lambda}+
\delta_{\gamma\delta}I_{\alpha\beta,\varkappa\lambda}),\nonumber\\
& {}^4v_{\mu\nu,\alpha\beta,\gamma\delta}=&
2q_\varkappa[I_{\varkappa\lambda,\alpha\beta}I_{\gamma\delta,\nu\lambda}(k-q)_\mu
+I_{\varkappa\lambda,\alpha\beta}I_{\gamma\delta,\mu\lambda}(k-q)_\nu
\nonumber\\
&&-I_{\varkappa\lambda,\gamma\delta}I_{\alpha\beta,\nu\lambda}
k_\mu -I_{\varkappa\lambda,\gamma\delta}I_{\alpha\beta,\mu\lambda}
k_\nu] \nonumber\\ &&+
q^2(I_{\lambda\mu,\alpha\beta}I_{\gamma\delta,\lambda\nu}
+I_{\lambda\nu,\alpha\beta}I_{\gamma\delta,\lambda\mu})+
\delta_{\mu\nu}q_\varkappa
q_\lambda(I_{\alpha\beta,\varkappa\rho}I_{\rho\lambda,\gamma\delta}
+I_{\gamma\delta,\varkappa\rho}I_{\rho\lambda,\alpha\beta}),\nonumber\\
& {}^5v_{\mu\nu,\alpha\beta,\gamma\delta}=&[k^2+(k-q)^2]\left(
I_{\lambda\mu,\alpha\beta}
I_{\gamma\delta,\lambda\nu}-\frac{1}{2}\,\delta_{\mu\nu}
P_{\alpha\beta,\gamma\delta}\right) \nonumber\\ && - k^2
\delta_{\gamma\delta}
I_{\mu\nu,\alpha\beta}-(k-q)^2\delta_{\alpha\beta}I_{\mu\nu,\gamma\delta}.\nonumber
\end{eqnarray}
In this vertex one can also neglect, to our accuracy, the last
structure ${}^5v_{\mu\nu,\alpha\beta,\gamma\delta}$.

\section{Universality of Spin-Independent Effects}

\begin{figure}[h]
\begin{center}
\includegraphics*{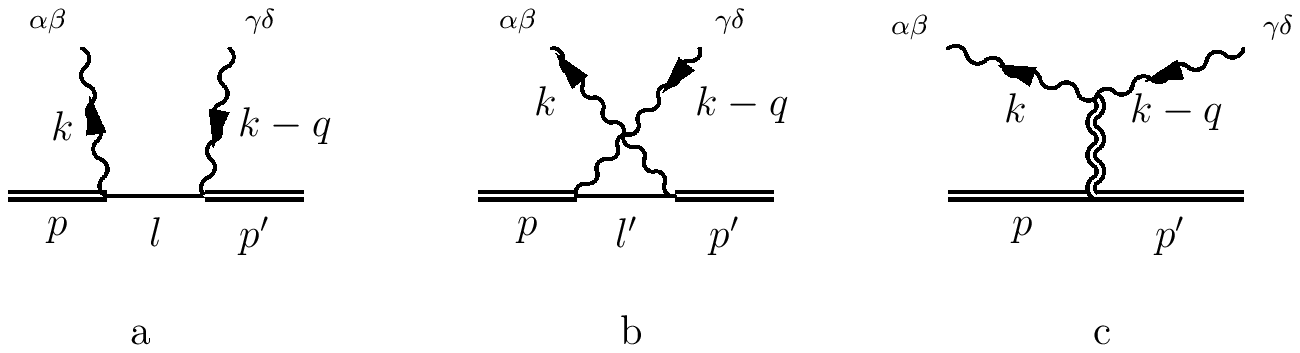}\hfill
\caption{Pole diagrams}\label{compton}
\end{center}
\end{figure}

Let us address at first the lowest-order $s$- and $u$-pole
diagrams for graviton scattering, presented in Fig.
\ref{compton}a,b.

We start with a scalar particle. The terms with the Kronecker
$\delta$ in the single-graviton vertices (\ref{1sc}) cancel here
the $s$- and $u$-pole denominators. It can be easily demonstrated
that the arising contact contributions combine in the sum of the
two diagrams into
\beq\label{cor_eff}
V^{(0)\prime}_{\alpha\beta,\,\gamma\delta}= i
\,\frac{\varkappa^2}{4}\,[\delta_{\alpha\beta}(p_\gamma
p{\prime}_\delta + p{\prime}_\gamma p_\delta) +
\delta_{\gamma\delta} (p_\alpha p{\prime}_\beta + p{\prime}_\alpha
p_\beta )]\,=\,i
\,\frac{\kappa^2}{4}\,(\delta_{\alpha\beta}T^{(0)}_{\gamma \delta}
 + \delta_{\gamma\delta}T^{(0)}_{\alpha\beta}).
\eeq
In the course of these transformations we omit the terms with
extra powers of the graviton momenta since after subsequent loop
integration they do not lead to $\ln q^2$ in the result. Combining
this induced term with (\ref{v2}), we arrive at the total
effective two-graviton vertex for a scalar particle:
\beq\label{0eff}
V_{\varkappa\lambda,\rho\sigma}^{(0)eff} =
i\varkappa^2\,I_{\varkappa\lambda,\alpha\delta}I_{\delta\beta,\rho\sigma}T^{(0)}_{\alpha\beta}
 =
i\,\frac{\varkappa^2}{2}\,I_{\varkappa\lambda,\alpha\delta}I_{\delta\beta,\rho\sigma}P_\alpha
P_\beta\,.
\eeq

For spinor particles the single-graviton vertices (\ref{1sp}) also
contain terms with the Kronecker~$\delta$. Proceeding here with
the $s$- and $u$-pole diagrams in the same way as in the scalar
case, we obtain the following correction to the two-graviton
vertex:
\beq\label{1/2cor_eff}
V^{(1/2)\prime}_{\alpha\beta,\,\gamma\delta}= \,i
\,\frac{\varkappa^2}{4}\,(\delta_{\alpha\beta}T^{(1/2)}_{\gamma
\delta}
 + \delta_{\gamma\delta}T^{(1/2)}_{\alpha\beta}).
\eeq
Then the total effective two-graviton vertex for a spinor particle
is
\beq\label{1/2eff}
V_{\varkappa\lambda,\rho\sigma}^{(1/2)eff} =
i\,\frac{3}{4}\,\varkappa^2\,I_{\varkappa\lambda,\alpha\delta}
I_{\delta\beta,\rho\sigma}T^{(1/2)}_{\alpha\beta}\,.
\eeq

If one is interested in spin-independent effects in the graviton
scattering off a spinor particle, one more step is possible. The
spinor structure of the numerators in the $s$- and $u$-pole
diagrams can be transformed as follows:
\beq
\bar{u}\,(p^{\prime})\gamma^\sigma(\hat{l}+m)\gamma^\omega u(p)=
\bar{u}\,(p^{\prime})\,[l^\sigma\gamma^\omega+l^\omega\gamma^\sigma-
(\hat{l}-m)\delta^{\sigma\omega}+i\gamma^5\epsilon^{\sigma\xi\omega\eta}
l_\xi \gamma_\eta+m \sigma_{\sigma\omega}]\,u(p).
\eeq
The term $\bar{u}\,(p^{\prime})(\hat{l}-m)u(p)$ in this
expression, when averaged over spins, transforms to $l^2-m^2$
(here we omit again a term proportional to $q^2$). After
cancelling the denominators, the sum of these terms in the $s$-
and $u$-pole diagrams reduces to
\beq\label{sav}
V^{(1/2)\prime\prime}_{\varkappa\lambda,\,\rho\sigma} =
\frac{i\,\varkappa^2}{8}\,I_{\varkappa\lambda,\,\mu\beta}
\,I_{\rho\sigma,\,\beta\alpha}P_\mu P_\alpha
\eeq
Since the spin-averaged energy-momentum tensor for spinors
coincides with the scalar one, which is equal to $P_\mu
P_\alpha/2$, the spin-independent term in the sum of
(\ref{1/2eff}) and (\ref{sav}) reduces to (\ref{0eff}). In other
words, we can single out from the fermion diagrams the sum of
structures that, after averaging over spins, coincides with the
effective sea-gull for a scalar particle.

At last, it can be easily demonstrated that all other terms in the
numerators of the $s$- and $u$-pole spinor diagrams, after
averaging over the spins, coincide with the required accuracy with
the corresponding terms in scalar diagrams.

As to the diagram Fig. \ref{compton},c, with the graviton pole in
the $t$-channel, here the coincidence between the scalar and
spin-averaged spinor cases is obvious.

To summarize, the sum of scalar and spin-averaged spinor tree
amplitudes, and hence the sum of the corresponding loop diagrams,
coincide with the required accuracy.

\section{Spin-Independent Effective Amplitudes}

We start the discussion of the loops with the vacuum polarization
diagrams, Fig. \ref{loop}. The covariant effective Lagrangian
corresponding to the sum of these loops was derived in \cite{hoo}
with dimensional regularization (see also \cite{dm}). It is
\beq\label{ld}
L_{RR} =\,-\,\frac{1}{960\pi^2(4-d)}\,\sqrt{-g} \left(42
R_{\mu\nu}R^{\mu\nu} + R^2 \right);
\eeq
here, as usual, $g$ is the determinant of the metric tensor,
$R_{\mu\nu}$ is the Ricci tensor, $R=R^\mu_\mu$.

For our purpose Lagrangian (\ref{ld}) is conveniently rewritten as
\cite{don}
\beq
L_{RR} =-\,\frac{1}{1920\pi^2}\, \ln |\,q^2|\, \left(42
R_{\mu\nu}R^{\mu\nu} +  R^2 \right);
\eeq
We will be interested in particular in the situation where at
least one of the particles is considered in the static limit. In
this case $|\,q^2|\rightarrow \q^2$, and in the coordinate
representation we obtain
\beq\label{rr} L_{RR}=\,\frac{1}{3840\pi^3 r^3}\,
\left(42 R_{\mu\nu}R^{\mu\nu} + R^2 \right).
\eeq
\begin{figure}
\hspace*{1.6cm}
\includegraphics[width=32mm]{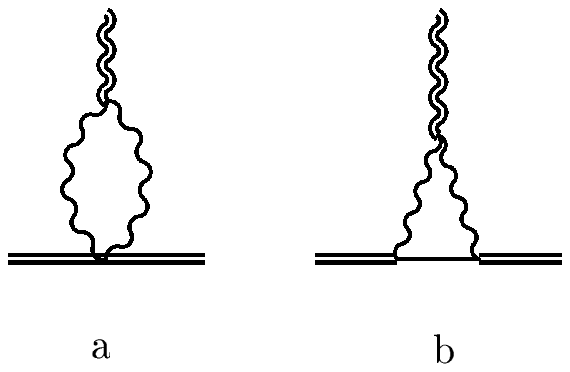}
\hspace{3.5cm}
\includegraphics[width=32mm]{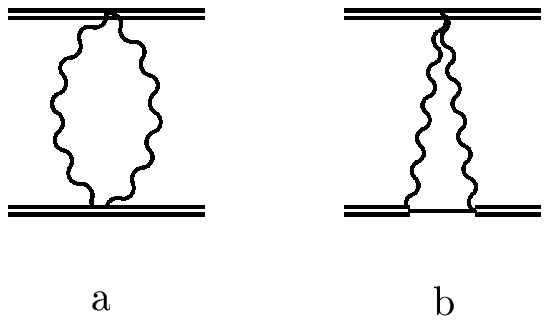}
\hspace{.35cm}
\includegraphics[width=32mm]{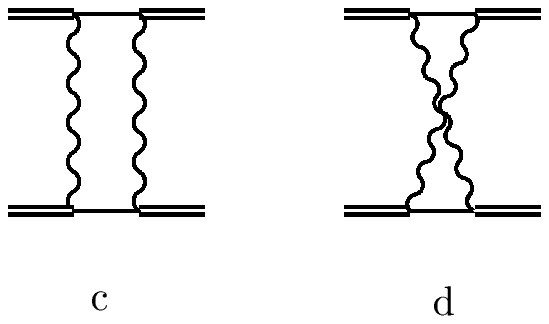}\\
\hspace*{1 cm}
\parbox[t]{.33\textwidth}{
\caption{Vertex diagrams} \label{vertex}}\hspace{2cm}
\parbox[t]{.45\textwidth}{\caption{Scattering
diagrams}\label{box}}\hfill
\end{figure}

The next set of diagrams, Fig. \ref{vertex}, refers to the vertex
part. The corresponding effective operator is
\beq\label{rt}
L_{RT} = -\,\frac{k}{8\pi^2 r^3} \left(3R_{\mu\nu}T^{\mu\nu} - 2 R
T \right); \quad T=T^\mu_\mu.
\eeq
Here and below $T^{\mu\nu}$ is the spin-independent part of the
total energy-momentum tensor of matter.

At last, diagrams in Fig. \ref{box}. The first two of them,
diagrams in Fig. \ref{box}a,b, as well as diagrams in Figs.
\ref{loop} and \ref{vertex}, depend only on the momentum transfer
$t=q^2$. As to the box diagrams in Fig. \ref{box}c,d, their
contribution is partly reducible to the same structure as that of
diagrams in Fig. \ref{box}a,b. The sum of all these $t$-dependent
effective operators originating from diagrams in Fig. \ref{box} is
\beq\label{tt}
L_{TT} = \,\frac{k^2}{\pi r^3}\, T^2.
\eeq

The irreducible contribution of the $s$-channel box diagram
\ref{box}c is
\[
M_s = \;\frac{k^2 [(s-m_1^2-m^2_2)^2 - 2m_1^2 m_2^2]^2}{m_1^2
m_2^2|\,q^2\,|}\;\ln\frac{|\,q^2\,|}{\la^2}
\]
\beq\label{s3}
\times \,\frac{1}{\sqrt{(s-m_-^2)(s-m_+^2)}}\,
\ln\frac{\sqrt{(s-m_-^2)}+\sqrt{(s-m_+^2)}}{\sqrt{(s-m_-^2)}-\sqrt{(s-m_+^2)}}\;.
\eeq
Here $m_1$, $m_2$ are the particle masses, $m_{\pm} = (m_1 \pm
m_2)$, $s~=~(p_1~+~p_2)^2$, $p_1$, $p_2$ are the incoming
4-momenta.

The irreducible contribution $M_u$ of the $u$-channel diagram in
Fig. \ref{box}d is obtained from formula (\ref{s3}) by the
substitution $s \to u~=~(p_1~-~p_2~-~q)^2$, with the corresponding
analytic continuation.

The expressions for $M_s$ and $M_u$ are convergent in the
ultraviolet sense, but diverge in the infrared limit, depending
logarithmically on the ``graviton mass'' $\lambda$. As usual, such
behaviour is directly related to the necessity to cancel the
infrared divergence in the Bremsstrahlung diagrams (of course, the
gravitational Bremsstrahlung in the present case). The box
diagrams \ref{box}c,d were considered previously by Donoghue,
Torma \cite{dot} from a different point of view.

As to the three Lagrangians (\ref{rr}), (\ref{rt}), (\ref{tt}), in
virtue of the Einstein equations
\beq\label{ein}
R_{\mu\nu}= 8\pi k \left(T_{\mu\nu}-
\frac{1}{2}g_{\mu\nu}T\right),
\eeq
they can be conveniently combined into
\beq\label{tot}
L_{tot} = -\,\frac{k^2}{60 \pi r^3}\,
\left(138\,T_{\mu\nu}T^{\mu\nu} - 31 T^2 \right).
\eeq

The irreducible amplitudes generated by the box diagrams
\ref{box}c,d depend nontrivially on $s$ and $u$, respectively, (in
line with their simple dependence on $\ln |q^2|/|q^2|$).
Therefore, they cannot be reduced to a product of energy-momentum
tensors.

\section{Quantum Corrections to Metric}

The effects due to Lagrangian (\ref{tot}) can be conveniently
interpreted as generated by quantum corrections to metric. To
obtain these corrections, let us split the total energy-momentum
tensor $T_{\mu\nu}$ into those of a static central body and of a
light probe particle, $T^o_{\mu\nu}$ and $t_{\mu\nu}$,
respectively. Then, by variation in $t^{\mu\nu}$ of the
expression, resulting in this way from (\ref{tot}), we obtain a
tensor which can be interpreted as a quantum correction
$h^{(q)}_{\mu\nu}$ to the metric created by the central body:
\beq\label{h}
h^{(q)}_{\mu\nu}= \frac{k^2}{15\pi r^3}\,\left(138 T^o_{\mu\nu} -
31 \de_{\mu\nu}T^o \right).
\eeq
It follows immediately from this expression that
\beq\label{h00}
h^{(q)}_{00}= \,\frac{107}{15}\,\frac{k^2}{\pi r^3}\, T^o_{00} =
\,\frac{107}{15}\,\frac{k^2 M}{\pi r^3}\,,
\eeq
where $M$ is the mass of the central body.

For the space components $h^{(q)}_{mn}$ of metric created by a
heavy body at rest, one might expect naively from formula
(\ref{h}) that they are
\[
\,\frac{31}{15}\,\frac{k^2}{\pi r^3}\,\de_{mn} T^o_{00} =
\,\frac{31}{15}\,\frac{k^2 M}{\pi r^3}\,\de_{mn}\,.
\]
However the calculation of $h^{(q)}_{mn}$ demands actually some
modification of formula~(\ref{h}). The point is that we work with
the gauge condition (\ref{harm}) for the graviton field. It is
only natural to require that the resulting effective field
$h^{(q)}_{mn}$ should satisfy the same condition which simplifies
now to $h^{(q)\mu}_{\quad\nu,\mu} - (1/2)
h^{(q)\mu}_{\quad\mu,\nu} =0$. Thus obtained space metric is
\beq\label{hmn}
h^{(q)}_{mn}= \,\frac{k^2 M}{\pi
r^3}\left\{\frac{31}{15}\,\de_{mn} -\,\frac{76}{15}\left[\frac{r_m
r_n}{r^2}\,+\ln\left(\frac{r}{r_0}\right)\left(\de_{mn}-3\,\frac{r_m
r_n}{r^2}\right)\right]\right\}.
\eeq

Technically, the expression in square brackets in (\ref{h00})
originates from the terms containing structures of the type
$\partial_\mu T^{\mu\nu}$. Generally speaking, they arise when
calculating Lagrangians (\ref{rt}), (\ref{tt}), and (\ref{tot}),
but are omitted therein since they vanish on-mass-shell. Thus
these terms are absent in (\ref{h}). But they can be restored by
rewriting, by means of the Einstein equations (\ref{ein}), the net
result (\ref{tot}) as
\beq\label{tot1}
L_{tot}=-\,\frac{1}{3840\pi^3 r^3}\, \left(138
R_{\mu\nu}R^{\mu\nu} - 31 R^2\right),
\eeq
and then attaching energy-momentum tensors to the double wavy
lines using the graviton propagators (\ref{gp}). The presence of
$\ln(r/r_0)$, where $r_0$ is some normalization point, is quite
natural here if one recalls $\ln |q^2|$ in the momentum
representation. Fortunately, this term in the square brackets does
not influence physical effects.

The obtained quantum corrections to metric $h^{(q)}_{00}$ and
$h^{(q)}_{mn}$ are universal, i. e. are the same when created by a
spinless or spinning heavy point-like particle.

Our results (\ref{h00}), (\ref{hmn}) differ from the corresponding
ones of \cite{bbo}.  The main reason is that the contribution of
operator (\ref{tt}) to metric is absent in \cite{bbo}. This
omission does not look logical to us: on-mass-shell one cannot
tell this operator from other ones (see (\ref{tot}),
(\ref{tot1})). One more disagreement is due perhaps to the same
inconsistency: the contribution of operator (\ref{rt}) to metric,
as given in \cite{bbo}, is two times smaller than ours.

In addition, the Fourier-transformation of $\; (q_m q_n/\q^2)\,
\ln \q^2\;$ is performed in \cite{bbo} incorrectly, which gives a
wrong result ($\,r_m r_n/r^2$ only) for the term in the square
brackets in (\ref{hmn}).

In conclusion of this section, let us consider the $0n$ component
of tensor (\ref{h}). It is
\beq\label{h0n}
h^{(q)}_{0n}= \,\frac{46}{5}\,\frac{k^2}{\pi r^3}\, T^o_{0n} =
\,-\,\frac{46}{5}\,\frac{k^2 M \bv}{\pi r^3}\,;
\eeq
here $\bv$ is the velocity of the source.

We are interested now in the situation corresponding to a compound
central body, rotating with the angular velocity $\vom$, but with
its centre of mass being at rest. Here the velocity of a separate
element of the body is $\bv = \vom \times \vrho$, where $\vrho$ is
the coordinate of this element. Besides, in formula (\ref{h0n})
one should shift $\br \to \br + \vrho$. Then, following \cite{ll},
\S 106, Problem 4, we obtain a quantum correction to the Kerr
metric:
\beq\label{hS}
h^{(q)}_{0n}= \,\frac{69}{5}\,\frac{k^2}{\pi r^5}\,[\bS \times
\br].
\eeq

Let us emphasize that here spin $\bS$ is in fact the internal
angular momentum of a rotating compound central body with spinless
constituents. We cannot see any reason why this last quantum
correction (\ref{hS}) should be universal (as distinct from
$h^{(q)}_{00}$ and $h^{(q)}_{mn}$). If instead of a compound body
discussed here, we deal with a particle of spin $1/2$, the general
structure of $h^{(q)}_{0n}$ is of course the same, but the
numerical coefficient can be quite different.

The last problem, that of a quantum correction to the Kerr metric
created by a particle of spin $1/2$, was addressed by
Bjerrum-Bohr, Donoghue, Holstein~\cite{bbo}. However, their
treatment of this correction causes the same objections: the
contribution of operator (\ref{tt}) to $h^{(q)}_{0n}$ is missed at
all, and the corresponding effect of operator (\ref{rt}) is not
taken into account properly.

\section{Quantum Corrections to Gravitational Effects I}

We start with the correction to the Newton law. As usual, it is
generated by the 00 component of metric. Here expression
(\ref{h00}) gives
\beq\label{Nr}
U^{qr}(r) = \,\frac{107}{30}\,\frac{k^2 M m}{\pi r^3}\,.
\eeq

However, now in line with (\ref{h00}), we should take into account
the irreducible contribution of the box diagrams \ref{box}c,d,
which cannot be reduced to metric. Having in mind other
applications, we write the sum of the two amplitudes, retaining in
it not only terms of zeroth order in $c^{-2}$, but of first order
as well:
\beq\label{irr}
M_{s} + M_{u} = - k^2 m_1 m_2 \ln (\q^2 -
\om^2)\,\frac{2}{3}\left(23 + \,\frac{524}{5}\;\frac{p_1 p_2 - m_1
m_2}{m_1 m_2}\,\right).
\eeq
In the static limit, $\om \to 0$, $p_1 p_2 \to m_1 m_2$,
expression (\ref{irr}) reduces to
\beq\label{irr0}
M_{s} + M_{u} \to -\,\frac{46}{3}\,k^2 m_1 m_2 \ln \q^2.
\eeq
Changing the sign (we are going over from amplitude to potential)
and performing the Fourier transformation, we obtain~\cite{kk,bb}
\beq\label{Nirr}
U^{qi}(r) = -\,\frac{23}{3}\,\frac{k^2 M m}{\pi r^3}\,.
\eeq
Thus, the net correction to the Newton law is
\beq
U^q(r) = -\,\frac{41}{10}\,\frac{k^2 M m}{\pi r^3}\,.
\eeq

This result was also cross-checked and confirmed by the
independent calculation in the usual harmonic gauge, with the
field variables $\psi^{\mu\nu}=\sqrt{-g}\,
g^{\mu\nu}-\de^{\mu\nu}$ and the gauge condition $\pa_\mu
\psi^{\mu\nu}= 0$.

Let us consider now the quantum correction to the interaction of
the orbital momentum $\bl$ of a light particle with its own spin
$\s$, i. e. to the gravitational spin-orbit interaction. It is
most easily obtained with the general expression for the frequency
$\vom$ of the spin precession in a gravitational field derived
in~\cite{kp}. For a nonrelativistic particle in a weak static
centrally-symmetric field this expression simplifies to
\beq
\omega_i =\,\frac{1}{2}\,\ep_{imn}(\ga_{mnk}v_k + \ga_{0n0} v_m).
\eeq
Here
\[
\ga_{mnk}=\,\frac{1}{2}\,(\partial_m h_{nk}-\partial_n h_{mk}),
\quad \ga_{0n0}=-\,\frac{1}{2}\,\partial_n h_{00}
\]
are the Ricci rotation coefficients, $\bv$ is the particle
velocity (the present sign convention for $\vom$ is opposite to
that of \cite{kp}). A simple calculation results in
\beq\label{ls}
U^q_{ls}(r) = -\,\frac{169}{20}\,\frac{k^2}{\pi
r^5}\,\frac{M}{m}\,(\bl \s).
\eeq

And finally, with formula (\ref{hS}) we derive easily the quantum
correction to the interaction of the orbital momentum $\bl$ of a
light particle with the internal angular momentum (spin) $\bS$ of
a compound central body, i. e. to the Lense-Thirring effect:
\beq\label{lt}
U^{q,r}_{LT}(r) = -\,\frac{69}{5}\,\frac{k^2}{\pi r^5}\,(\bl \bS).
\eeq

\section{Aside on Classical Relativistic Corrections}

In this section we consider at first the classical
velocity-dependent correction to the Newton law. On the one hand
this is an introduction to the derivation in the next section of
quantum velocity-dependent corrections. On the other hand, this is
necessary for the discussion of another, velocity-independent
relativistic correction to the Newton law. The derivation of the
classical velocity-independent correction in the diagram technique
served in \cite{kk,bb} as a check of calculations of quantum
corrections to the Newton law.

Let us consider the Born scattering amplitude with the graviton
exchange in the harmonic gauge:
\beq\label{B}
M_B =\, 8\pi k\,\frac{T^1_{\mu\nu}\,T^2_{\mu\nu} - 1/2\,
T^1_{\mu\mu}\,T^2_{\nu\nu}}{\q^2 - \om^2}\;.
\eeq
Here $T^{1,2}_{\mu\nu}$ are the energy-momentum tensors of
particles with masses $m_{1,2}$ and velocities $\bv_{1,2}$,
respectively. To the adopted accuracy, the numerator simplifies to
\[
\frac{1}{2}\; T^1_{00}\,T^2_{00} -  2 T^1_{0n}\,T^2_{0n} =
\;\frac{m_1 m_2}{2}\,(1 - 4\,\bv_1 \bv_2).
\]
Then we expand the denominator to first order in $\om^2/\q^2$, and
thus arrive at the following expression
\[
\frac{4\pi k m_1 m_2}{\q^2}\,\left(1 - 4\,\bv_1
\bv_2+\,\frac{\om^2}{\q^2}\right)\,.
\]
The term of zeroth order in $c^{-2}$ in this formula, $4\pi k m_1
m_2/\q^2$, is obviously (after the necessary sign reversal) the
Fourier-transform of the Newton potential. However, we are
interested here in the terms of first order in $c^{-2}$. To
transform $\om^2/\q^2$, let us note that $\om$ is in fact the
energy difference between the initial and final energies of a
particle. The particles can be considered now as nonrelativistic,
so this difference transforms (to first order in $\p^{\prime} - \p
$) as follows:
\[
\varepsilon^{\prime} - \varepsilon = (\p^{\prime} - \p)\bv.
\]
Therefore, the terms of first order in $c^{-2}$ are rewritten as
\[
\frac{4\pi k m_1 m_2}{\q^2}\,\left[ - 4\,\bv_1 \bv_2+\,\frac{(\q
\bv_1)(\q \bv_2)}{\q^2}\right]\,.
\]
The Fourier-transform of this expression, taken with the opposite
sign, is the well-known relativistic velocity-dependent correction
to the Newton potential~\cite{eih,ec,ll}:
\beq\label{cv}
U^{cl}_{vv} =\,\frac{ k m_1 m_2}{2 r}\,[ 7 \bv_1 \bv_2 + (\n
\bv_1)(\n \bv_2)], \quad \n=\,\frac{\br}{r}\,.
\eeq
We follow here essentially the derivation by Iwasaki~\cite{iw}.

At least as simple is the derivation of the relativistic
velocity-independent correction to the Newton potential. In the
harmonic gauge the metric created by a point-like mass $m_1$ is
\beq
ds^2 =\,\frac{r-k m_1}{r+k m_1}\,dt^2 - \,\frac{r+k m_1}{r-k
m_1}\,dr^2 - (r+k m_1)^2 (d\theta^2 + \sin^2 \theta d\phi^2).
\eeq
In the expansion in $r_g$ of the classical action $-m_2\int ds$
for a probe particle of mass $m_2$, the second-order term is $-k^2
m_1^2 m_2/2r^2$. Now, reversing the sign (to go over from a
Lagrangian to a potential) and restoring the symmetry between
$m_1$ and $m_2$, we arrive at the discussed correction:
\beq\label{cl}
U^{cl} =\,\frac{ k^2 m_1 m_2 (m_1 + m_2)}{2 r^2}\,.
\eeq

The classical correction (\ref{cl}) was found long ago by
Einstein, Infeld, Hoffmann~\cite{eih}, Eddington, Clark~\cite{ec}
(see also the textbook~\cite{ll}, \S 106), and derived later
in~\cite{iw} by calculating in the harmonic gauge the
corresponding parts of diagrams \ref{vertex}b, and \ref{box}b,c,d.
A subtle point of the last calculation~\cite{iw} refers to the box
diagrams \ref{box}c,d. Obviously, the classical $c^{-2}$
contribution of these diagrams contains in particular the result
of iteration of the usual Newton interaction and the
velocity-dependent interaction (\ref{cv}). Therefore, the result
of this iteration should be subtracted from the sum of the
contributions of diagrams \ref{vertex}b, \ref{box}b,c,d. This has
been done properly by Iwasaki~\cite{iw}).

However, Bjerrum-Bohr, Donoghue, and Holstein argue (see section
2.1 in~\cite{bb}) that in the scattering problem, as distinct from
the bound-state one, this subtraction is unnecessary. They claim
that there is a difference between what they call ``the lowest
order scattering potential'' without this subtraction, and the
classical correction $U^{cl}$ which they call the bound state
potential. For our part, we do not see any difference of principle
between the bound state problem and the scattering
one\footnote{For instance, the second Born approximation to a
scattering amplitude is as legitimate notion, as the second-order
correction to a bound state energy.}, and thus believe that it is
just (\ref{cl}) which should be considered as the relativistic
correction to the Newton law, both in the scattering and bound
state problems.

\section{Quantum Corrections to Gravitational Effects II}

We address now the quantum correction to the classical
velocity-dependent gravitational interaction (\ref{cv}). We start
with the amplitude (\ref{tot}) written in the momentum
representation:
\beq\label{totq}
L_{tot} = \,\frac{k^2}{30}\, \ln |q^2|
\left(138\,T_{\mu\nu}T^{\mu\nu} - 31 T^2 \right).
\eeq
As distinct from the previous quantum corrections, here we go
beyond the static approximation, and in the spirit of  the
previous section expand $\ln |q^2| = \ln (\q^2-\omega^2)$ to first
order in $\omega^2$. Following further the same lines of
reasoning, we arrive easily at the quantum velocity-dependent
correction:
\beq\label{vv}
U^{q,r}_{vv}(\br) = -\,\frac{k^2 m_1 m_2}{60 \pi r^3}\;[445 (\bv_1
\bv_2) + 321 (\n\bv_1)(\n \bv_2) ], \quad \n=\,\frac{\br}{r}\,.
\eeq

With formula (\ref{vv}) we can derive (in the spirit of \cite{ll},
\S 106, Problem 4) the quantum correction to the spin-spin
interaction of compound bodies 1 and 2 rotating with the angular
velocities $\vom_1$ and $\vom_2$, but their centres of masses
being at rest. Here the velocity of a separate element of the body
$i$ is $\bv_i= \vom_i\times \vrho_i$, where $\vrho_i$ is the
coordinate of this element counted off the center of mass of this
body. Then in formula (\ref{vv}), where $\br = \br_1 -\br_2$, we
shift $\br \to \br  + \vrho_1 - \vrho_2$. Following again
\cite{ll}, \S 106, Problem 4, we obtain in this way
\beq\label{ss}
U^{q,r}_{ss}(\br) = \,\frac{69}{10}\,\frac{k^2}{\pi r^5}\,[3(\bS_1
\bS_2)-5(\n \bS_1)(\n \bS_2)], \quad \n=\,\frac{\br}{r}\,.
\eeq
Here $\bS_i$ are the internal angular momenta (spins) of the
rotating compound central bodies.

Let us note that in the same way one can derive also the quantum
correction (\ref{lt})  to the Lense-Thirring effect.

At last, let us consider the corresponding corrections induced by
the irreducible amplitude~(\ref{irr}) which is conveniently
rewritten now as
\beq\label{irr1}
M_{s} + M_{u} = - k^2 m_1 m_2 \ln (\q^2 -
\om^2)\,\frac{2}{3}\left(23 - \,\frac{524}{5}\, \bv_1
\bv_2\right).
\eeq
This amplitude also generates quantum corrections to the
velocity-dependent, Lense-Thirring, and spin-spin interactions.
The calculations are practically the same as the previous ones,
and give, respectively:
\begin{equation}\label{vvirr}
U^{q,\,irr}_{vv}({\mathbf r}) = \,\frac{k^2 m_1 m_2}{10 \pi
r^3}\;[311 ({\mathbf v}_1 {\mathbf v}_2) + 115 ({\mathbf
n}{\mathbf v}_1)({\mathbf n} {\mathbf v}_2) ],
\end{equation}
\begin{equation}
U^{q,\,irr}_{LT}(r) = \,\frac{262}{5}\,\frac{k^2}{\pi
r^5}\,({\mathbf l} {\mathbf S}),
\end{equation}
\begin{equation}
U^{q,\,irr}_{ss}({\mathbf r}) = -\,\frac{131}{5}\,\frac{k^2}{\pi
r^5}\,[3({\bS}_1 {\bS}_2) -5({\mathbf n} {\bS}_1)({\mathbf n}
{\bS}_2)].
\end{equation}

Now, combining these contributions with those originating from
quantum corrections to metric, we obtain finally
\begin{equation}
U^{q}_{vv}({\mathbf r}) = U^{q,r}_{vv} ({\mathbf r}) +
U^{q,\,irr}_{vv}({\mathbf r}) = \,\frac{k^2 m_1 m_2}{60 \pi
r^3}\;[1421 ({\mathbf v}_1 {\mathbf v}_2) + 369 ({\mathbf
n}{\mathbf v}_1)({\mathbf n} {\mathbf v}_2) ],
\end{equation}
\begin{equation}
U^{q}_{LT}(r) = U^{q,r}_{LT}(r) + U^{q,\,irr}_{LT}(r) =
\,\frac{193}{5}\,\frac{k^2}{\pi r^5}\,({\mathbf l} {\mathbf S}),
\end{equation}
\begin{equation}
U^{q}_{ss}({\mathbf r}) = U^{q,r}_{ss} ({\mathbf r}) +
U^{q,\,irr}_{ss}({\mathbf r}) = -\,\frac{193}{10}\,\frac{k^2}{\pi
r^5}\,[3({\bS}_1 {\bS}_2) -5({\mathbf n} {\bS}_1)({\mathbf n}
{\bS}_2)].
\end{equation}

\subsection*{Acknowledgements}

We are grateful to N.G. Ural'tsev and A.I. Vainshtein for useful
discussions. The investigation was supported by the Russian
Foundation for Basic Research through Grant No. 03-02-17612.

\end{document}